# A Mass-in-Mass Chain and the Generalization of the Dirac Equation with an Eight-Component Wave Function and with Optical and Acoustic Branches of the Dispersion Relation

V. O. Turin (https://orcid.org/ 0000-0003-0450-2481) [a,*], Y. V. Ilyushina [b], P. A. Andreev [a], A. Yu. Cherepkova [a], D. D. Kireev [a], and I. V. Nazritsky [a]

[a] *Orel State University named after I.S. Turgenev, Orel, 302026 Russia*

[b] *Moscow Aviation Institute (National Research University), Moscow, 125993 Russia*

*e-mail: voturin@mail.ru

**Abstract**— The paper considers a slightly modified one-dimensional infinite mass-in-mass chain. In the case of the long-wave approximation, which corresponds to the transition to a continuous medium, we obtained a system of two equations, which is a generalization of the classical mechanics Klein-Gordon-Fock equation and has both optical and acoustic branches of the dispersion relation. Based on this classical mechanics system of equations, we have proposed a system of two relativistic quantum mechanics equations, which is a generalization of the relativistic quantum mechanics Klein-Gordon-Fock equation. Next, based on this system and following the Dirac approach, we have proposed the generalization of the Dirac equation for a free electron with an eight-component wave function in the form of a system of eight linear partial differential equations of the first order. Unlike the Dirac equation with a four-component wave function, which has only an optical branch of the dispersion relation, the generalized Dirac equation has both optical and acoustic branches of the dispersion relation, each of which has two branches with positive and negative energies, respectively. We have calculated phase and group velocities for all cases. For the positive and negative acoustic branches, the phase and group velocities are equal in modulus to the speed of light. For the positive and negative optical branches, the phase and group velocities have a structure like that of de Broglie waves. In the one-dimensional case, eight linearly independent solutions corresponding to eight combinations of two branches of dispersion, two signs of total energy, and two possible directions of spin orientation, each in the form of four plane waves, are obtained.

`

## INTRODUCTION

In [1, 2] a one-dimensional infinite mass-in-mass chain is considered (see Fig. 1a). This system is the simplest mechanical filter and implements the concept of effective mass. First proposed and investigated in 1898, this system continues to arouse interest at the present stage. For example, in [3, 4] the mass-in-mass chain was investigated in detail. In [3] a system of partial differential equations is derived for the case of a long-wave approximation, which corresponds to the transition to a continuous medium. Note that in the case of $M \gg m$, we can assume that the mass $m$ is attached by a spring with the spring constant $K$ to a stationary equilibrium position marked with a cross (see Fig. 1b). In our work, we consider a slightly modified system (see Fig. 1c). We introduced a harmonic interaction between loads of the same mass $M$ by adding springs with the spring constant $J$.

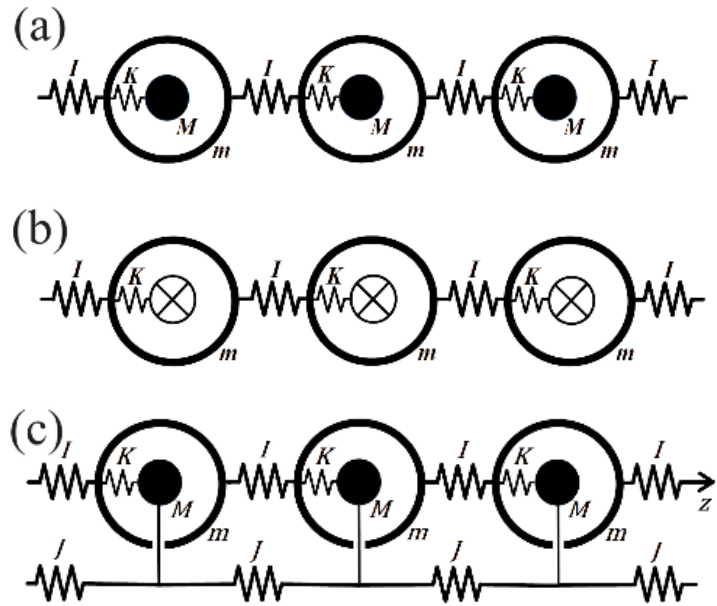

**Fig. 1.** One-dimensional infinite Vincent mass-in-mass chain (a), the Vincent mass-in-mass chain in the case of $M \gg m$ (the equilibrium position of the load $m$ is stationary and marked with a cross), and a modified mass-in-mass chain with the addition of harmonic interaction between loads with the same mass $M$.

The force of interaction between adjacent loads with different masses is proportional to the difference in their small displacements, with a proportionality coefficient $K$. The force of interaction between the nearest loads with the same mass $m$, with their small displacements from the equilibrium position $u_n$, is proportional to the difference of their displacements with the proportionality coefficient $I$. And the force of interaction between the nearest loads with the same mass $M$, with their small displacements from the equilibrium position $U_n$, is proportional to the difference of their displacements with the proportionality coefficient $J$. The coordinate of the equilibrium position of the $n$-th mass $M$ is denoted by $z_n = an$. In this case, the coordinate of the equilibrium position of the $n$-th mass $m$ coincides with the coordinate of the equilibrium position of the $n$-th mass $M$. Let us write Newton's second law for $n$-th masses $M$ and $m$. We get a system of equations:

$$\begin{cases} m \frac{d^2 u_n}{dt^2} = K(U_n - u_n) + I(u_{n-1} + u_{n+1} - 2u_n) \\ M \frac{d^2 U_n}{dt^2} = K(u_n - U_n) + J(U_{n-1} + U_{n+1} - 2U_n) \end{cases}. \quad (1)$$

We introduce characteristic frequencies and characteristic speeds:

$$\omega_O = \sqrt{\frac{K}{m}}, \quad \omega_A = \sqrt{\frac{K}{M}}, \quad \omega_m = \sqrt{\frac{I}{m}}, \quad \omega_M = \sqrt{\frac{J}{M}}, \quad s_m = a\omega_m, \quad s_M = a\omega_M. \quad (2)$$

Consider only long-wave oscillations, that is, oscillations with a wavelength $\lambda$ much longer than the period of the chain $a$ ($\lambda = 2\pi/k \gg a$). We obtain a system of equations for the corresponding one-dimensional continuous medium:

$$\begin{cases} \frac{\partial^2 u}{\partial t^2} = s_m^2 \frac{\partial^2 u}{\partial z^2} - \omega_O^2 (u - U) \\ \frac{\partial^2 U}{\partial t^2} = s_M^2 \frac{\partial^2 U}{\partial z^2} - \omega_A^2 (U - u) \end{cases}. \quad (3)$$

In the case of $M \gg m$ (or $\omega_A \ll \omega_O$) (see Fig. 1b), instead of the system of equations (3), the classical physics Klein-Gordon-Fock equation (KGF) is obtained [5-9]:

$$\frac{\partial^2 u}{\partial t^2} = s_m^2 \frac{\partial^2 u}{\partial z^2} - \omega_O^2 u. \quad (4)$$

Accordingly, we can conclude that the system of equations (3) is a generalization of the KGF equation (4). But the KGF equation is also an equation of relativistic quantum mechanics and, in the one-dimensional case of a free particle, has the form [10 - 12]:

$$\frac{\partial^2 \Psi}{\partial t^2} = c^2 \frac{\partial^2 \Psi}{\partial x^2} - \left(\frac{m_e c^2}{\hbar}\right)^2 \Psi. \quad (5)$$

This equation is obtained from (4) by replacing $u \to \Psi$, $s_m \to c$ ($c$ is the speed of light) and $\omega_O = \sqrt{K/m} \to m_e c^2/\hbar$. It is known that a number of problems are associated with the KGF equation in relativistic quantum mechanics, and a further generalization is the Dirac equation [13, 14], which solves these problems and was obtained by Dirac in 1928 using the linearization of the Hamiltonian from the KGF equation. The Dirac equation is a relativistically invariant equation of motion for the bispinor electron field, which is a four-component complex wave function.

## THE GENERALIZATION OF THE DIRAC EQUATION

By making in the system (3) a replacement

$$u \to \Psi, \quad s_m \to c, \quad s_M \to c, \quad \omega_O = \sqrt{K/m} \to m_e c^2/\hbar, \quad \omega_A = \sqrt{K/M} \to m_f c^2/\hbar, \quad (6)$$

similar to what we did before in the case of KGF equation (4), we obtain a system that is a generalization of the relativistic quantum mechanics one-dimensional KGF equation (5):

$$\begin{cases} \frac{\partial^2 \Psi}{\partial t^2} = c^2 \frac{\partial^2 \Psi}{\partial z^2} - \left(\frac{m_e c^2}{\hbar}\right)^2 (\Psi - \Phi) \\ \frac{\partial^2 \Phi}{\partial t^2} = c^2 \frac{\partial^2 \Phi}{\partial z^2} - \left(\frac{m_f c^2}{\hbar}\right)^2 (\Phi - \Psi) \end{cases}. \quad (7)$$

Accordingly, the question arises about the formal possibility of obtaining, on the basis of this system of second-order partial differential equations, which is a generalization of the KGF equation, a system of first-order partial differential equations, which is a generalization of the Dirac equation, similar to how the Dirac equation was obtained on the basis of the KGF equation. This paper is devoted to the answer to this question. Note that the detailed presentation there is in [15, 16] in Russian.

Here, we present the Dirac system of equations in the one-dimensional case for the projection of spin $\hbar/2$ on the $z$-axis:

$$\begin{cases} i\hbar \frac{\partial \Psi_1}{\partial t} = -i\hbar c \frac{\partial \Psi_3}{\partial z} + m_e c^2 \Psi_1 \\ i\hbar \frac{\partial \Psi_3}{\partial t} = -i\hbar c \frac{\partial \Psi_1}{\partial z} - m_e c^2 \Psi_3 \end{cases}. \quad (8)$$

We found that the following system of first-order equations is a generalization of the Dirac system of equations (8) in the one-dimensional case for the projection of spin $\hbar/2$ on the $z$-axis:

$$\begin{cases} i\hbar \frac{\partial \Psi_1}{\partial t} = -i\hbar c \frac{\partial \Psi_3}{\partial z} + \frac{m_e c^2}{\sqrt{1+\varepsilon^2}} (\Psi_1 - \Phi_1) \\ i\hbar \frac{\partial \Psi_3}{\partial t} = -i\hbar c \frac{\partial \Psi_1}{\partial z} - \frac{m_e c^2}{\sqrt{1+\varepsilon^2}} (\Psi_3 - \Phi_3) \\ i\hbar \frac{\partial \Phi_1}{\partial t} = -i\hbar c \frac{\partial \Phi_3}{\partial z} + \frac{m_f c^2}{\sqrt{1+\varepsilon^{-2}}} (\Phi_1 - \Psi_1) \\ i\hbar \frac{\partial \Phi_3}{\partial t} = -i\hbar c \frac{\partial \Phi_1}{\partial z} - \frac{m_f c^2}{\sqrt{1+\varepsilon^{-2}}} (\Phi_3 - \Psi_3) \end{cases}. \quad (9)$$

Here, we use the notation:

$$\varepsilon = m_f/m_e = \sqrt{m/M}. \quad (10)$$

To obtain the dispersion relation for system (9), we used a simple plane waves ansatz:

$$\Psi_1 = b_1 e^{-i\left(\frac{E}{\hbar}t - \frac{p_z}{\hbar}z\right)}, \; \Psi_3 = b_3 e^{-i\left(\frac{E}{\hbar}t - \frac{p_z}{\hbar}z\right)}, \; \Phi_1 = d_1 e^{-i\left(\frac{E}{\hbar}t - \frac{p_z}{\hbar}z\right)}, \; \Phi_3 = d_3 e^{-i\left(\frac{E}{\hbar}t - \frac{p_z}{\hbar}z\right)}. \quad (11)$$

The obtained homogeneous system of linear equations has a solution if its determinant is zero:

$$(E^2 - c^2 p_z^2)[E^2 - c^2 p_z^2 - (1 + \varepsilon^2) m_e^2 c^4] = 0. \quad (12)$$

This equation splits into two equations. An acoustic branch of the dispersion relation looks like this:

$$E^2 = c^2 p_z^2 \quad \text{and} \quad \Omega_A^2 = c^2 k_z^2. \quad (13)$$

An optical branch of the dispersion relation looks like this:

$$E^2 = c^2 p_z^2 + (1 + \varepsilon^2) m_e^2 c^4 \quad \text{and} \quad \Omega_O^2 = \omega_0^2 + \omega_A^2 + c^2 k_z^2 + c^2 k_z^2. \quad (14)$$

Consider the first pair of roots of (12). From equation (13), two linear (acoustic) dispersion relations follow:

$$E = \pm c\, p_z \,. \quad (15)$$

In the case $E > 0$ (sign (+) in equation (15)):

$$E = c\, p_z \quad \text{or} \quad \Omega_{A+} = c k_z. \quad (16)$$

Phase and group velocities:

$$v_{pA+} = \frac{\Omega_{A+}}{k_z} = c \quad \text{and} \quad v_{gA+} = \frac{d\Omega_{A+}}{dk_z} = c. \quad (17)$$

The corresponding amplitudes of waves (11):

$$b_{1A+} = b_{3A+} = d_{1A+} = d_{3A+}. \quad (18)$$

For the corresponding plane waves, we have the following:

$$\Psi_{1A+} = \Psi_{3A+} = \Phi_{1A+} = \Phi_{3A+} = b_{1A+} e^{-i(ct - z)\frac{p_z}{\hbar}} \,. \quad (19)$$

We can assume that the in-phase waves $\Psi_{1A+}$ and $\Psi_{3A+}$ are summed to form a total solution with spin $\hbar$, which resembles an electromagnetic wave, the quantum of which is a photon, the calibration boson of electromagnetic interaction, in which an electron and a positron participate. In the same way, we can assume that the in-phase waves $\Phi_{1A+}$ and $\Phi_{3A+}$ are summed, also forming a total solution with spin $\hbar$. It can be assumed that the resulting boson is somehow connected with particle-antiparticle pairs corresponding to the waves $\Phi_{1O}$ and $\Phi_{3O}$, which will be discussed below.

In the case $E < 0$ (sign (-) in equation (15)):

$$E = -|E| = -c\, p_z \quad \text{и} \quad \Omega_{A-} = -c k_z. \quad (20)$$

Phase velocity and group velocities:

$$v_{pA-} = \frac{\Omega_{A-}}{k_z} = -c. \quad v_{gA-} = \frac{d\Omega_{A-}}{dk_z} = -c. \quad (21)$$

The corresponding amplitudes of waves (11):

$$b_{1A-} = -b_{3A-} = d_{1A-} = -d_{3A-}. \quad (22)$$

For the corresponding plane waves, we have the following:

$$\Psi_{1A-} = \Phi_{1A-} = b_{1A-}e^{i(ct+z)\frac{p_z}{\hbar}}, \qquad \Psi_{3A-} = \Phi_{3A-} = -b_{1A-}e^{i(ct+z)\frac{p_z}{\hbar}}. \quad (23)$$

Note that:

$$\Psi_{1A-} + \Psi_{3A-} \equiv 0 \quad \text{and} \quad \Phi_{1A-} + \Phi_{3A-} \equiv 0. \quad (24)$$

That is, the waves are mutually compensating for each other. And we can assume that the observation of such waves is difficult.

Consider the second pair of roots of equation (12). From equation (14), two nonlinear (optical) dispersion relations follow:

$$E = \pm\sqrt{c^2 p_z^2 + m_e^2 c^4 (1+\varepsilon^2)}. \quad (25)$$

In case $E > 0$ (sign (+) in equation (25)):

$$E = \sqrt{c^2 p_z^2 + m_e^2 c^4 (1+\varepsilon^2)} \quad \text{and} \quad \Omega_{O+} = \sqrt{c^2 k_z^2 + \omega_O^2 + \omega_A^2}. \quad (26)$$

Phase and group velocities:

$$v_{pO+} = \frac{\Omega_{O+}}{k_z} = \sqrt{c^2 + \frac{\omega_O^2 + \omega_A^2}{k_z^2}} \quad \text{and} \quad v_{gO+} = \frac{d\Omega_{O+}}{dk_z} = \frac{c^2 k_z}{\Omega_{O+}} = \frac{c^2}{\sqrt{c^2 + \frac{\omega_O^2 + \omega_A^2}{k_z^2}}}. \quad (27)$$

The corresponding amplitudes of waves (11):

$$b_{3O+} = \frac{c\, p_z}{\sqrt{c^2 p_z^2 + m_e^2 c^4 (1+\varepsilon^2)} + m_e c^2 \sqrt{1+\varepsilon^2}} b_{1O+}. \quad (28)$$

$$d_{1O+} = -\varepsilon^2 b_{1O+}. \quad (29)$$

$$d_{3O+} = \frac{c\, p_z}{\sqrt{c^2 p_z^2 + m_e^2 c^4 (1+\varepsilon^2)} + m_e c^2 \sqrt{1+\varepsilon^2}} d_{1O+}. \quad (30)$$

For the corresponding plane waves, we have the following:

$$\Psi_{1O+} = b_{1O+} e^{-i\left(\frac{\sqrt{c^2 p_z^2 + m_e^2 c^4 (1+\varepsilon^2)}}{\hbar} t - \frac{p_z}{\hbar} z\right)}. \quad (31)$$

$$\Psi_{3O+} = \frac{c\, p_z}{\sqrt{c^2 p_z^2 + m_e^2 c^4 (1+\varepsilon^2)} + m_e c^2 \sqrt{1+\varepsilon^2}} b_{1O+} e^{-i\left(\frac{\sqrt{c^2 p_z^2 + m_e^2 c^4 (1+\varepsilon^2)}}{\hbar} t - \frac{p_z}{\hbar} z\right)}. \quad (32)$$

$$\Phi_{1O+} = -\varepsilon^2 b_{1O+} e^{-i\left(\frac{\sqrt{c^2 p_z^2 + m_e^2 c^4 (1+\varepsilon^2)}}{\hbar} t - \frac{p_z}{\hbar} z\right)}. \quad (33)$$

$$\Phi_{3O+} = -\varepsilon^2 \frac{c\, p_z}{\sqrt{c^2 p_z^2 + m_e^2 c^4 (1+\varepsilon^2)} + m_e c^2 \sqrt{1+\varepsilon^2}} b_{1O+} e^{-i\left(\frac{\sqrt{c^2 p_z^2 + m_e^2 c^4 (1+\varepsilon^2)}}{\hbar} t - \frac{p_z}{\hbar} z\right)}. \quad (34)$$

In case $E < 0$ (sign (-) in equation (25)):

$$E = -|E| = -\sqrt{c^2 p_z^2 + m_e^2 c^4(1+\varepsilon^2)} \quad \text{and} \quad \Omega_{0-} = -\sqrt{c^2 k_z^2 + \omega_0^2 + \omega_A^2}. \quad (35)$$

Phase and group velocities:

$$v_{p0-} = \frac{\Omega_{0-}}{k_z} = -\sqrt{c^2 + \frac{\omega_0^2 + \omega_A^2}{k_z^2}} \quad \text{and} \quad v_{g0-} = \frac{d\Omega_{0-}}{dk_z} = \frac{c^2 k_z}{\Omega_{0-}} = -\frac{c^2}{\sqrt{c^2 + \frac{\omega_0^2 + \omega_A^2}{k_z^2}}}. \quad (36)$$

The corresponding amplitudes of waves (11):

$$b_{30-} = -\frac{c\, p_z}{\sqrt{c^2 p_z^2 + m_e^2 c^4(1+\varepsilon^2)} - m_e c^2 \sqrt{1+\varepsilon^2}} b_{10-}. \quad (37)$$

$$d_{10-} = -\varepsilon^2 b_{10-}. \quad (38)$$

$$d_{30-} = -\frac{c\, p_z}{\sqrt{c^2 p_z^2 + m_e^2 c^4(1+\varepsilon^2)} - m_e c^2 \sqrt{1+\varepsilon^2}} d_{10-}. \quad (39)$$

For the corresponding plane waves, we have the following:

$$\Psi_{10-} = b_{10-} e^{-i\left(-\frac{\sqrt{c^2 p_z^2 + m_e^2 c^4(1+\varepsilon^2)}}{\hbar} t - \frac{p_z}{\hbar} z\right)}. \quad (40)$$

$$\Psi_{30-} = -\frac{c\, p_z}{\sqrt{c^2 p_z^2 + m_e^2 c^4(1+\varepsilon^2)} - m_e c^2 \sqrt{1+\varepsilon^2}} b_{10-} e^{-i\left(-\frac{\sqrt{c^2 p_z^2 + m_e^2 c^4(1+\varepsilon^2)}}{\hbar} t - \frac{p_z}{\hbar} z\right)}. \quad (41)$$

$$\Phi_{10-} = -\varepsilon^2 b_{10-} e^{-i\left(-\frac{\sqrt{c^2 p_z^2 + m_e^2 c^4(1+\varepsilon^2)}}{\hbar} t - \frac{p_z}{\hbar} z\right)}. \quad (42)$$

$$\Phi_{30-} = \varepsilon^2 \frac{c\, p_z}{\sqrt{c^2 p_z^2 + m_e^2 c^4(1+\varepsilon^2)} - m_e c^2 \sqrt{1+\varepsilon^2}} b_{10-} e^{-i\left(-\frac{\sqrt{c^2 p_z^2 + m_e^2 c^4(1+\varepsilon^2)}}{\hbar} t - \frac{p_z}{\hbar} z\right)}. \quad (43)$$

All branches of the dispersion relation are shown in Fig. 2. In all the cases considered, the condition known for electromagnetic waves in vacuum and for de Broglie waves is fulfilled:

$$v_p v_g = c^2. \quad (44)$$

Definitely, the waves $\Psi_{10}$ and $\Psi_{30}$ correspond to the particle - electron and antiparticle - positron. Interpreting the physical nature of the waves $\Phi_{10}$ and $\Phi_{30}$ is more difficult. Obviously, $\Phi_{10}$ corresponds to a particle, and $\Phi_{30}$ to an antiparticle. Because we assume that the corresponding rest mass $m_f$ is much smaller than the rest mass of an electron $m_e$, i.e. $m_f \ll m_e$, we will conditionally call them, by analogy with neutrino, "electrino" and "positrino". Note that neutrinos have similar properties: fundamental particles with half-integer spin $\hbar/2$ and a mass less than 0.12 eV (the mass of an electron is approximately 0.51 MeV, that is, almost four million times more). Because our study is related to the Dirac equation describing the electron and positron, that is, apparently, the electron neutrino and antineutrino may be candidates for the role of particles corresponding to the waves $\Phi_{10}$ and $\Phi_{30}$.

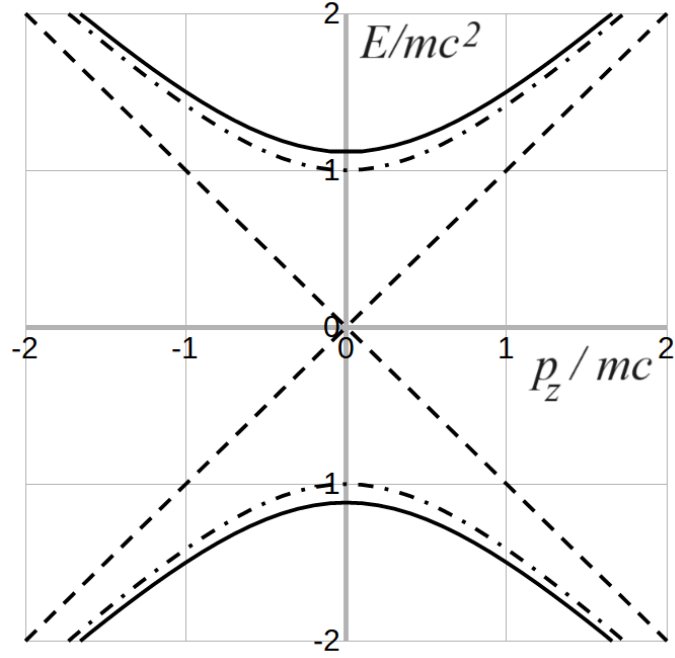

**Fig. 2.** All branches of the dispersion relation are presented. Dashed lines for the linear case of dispersion $E = \pm c\, p_z$. Solid lines for the nonlinear case of dispersion $E = \pm\sqrt{c^2 p_z^2 + m_e^2 c^4(1+\varepsilon^2)}$ with parameter $\varepsilon = 0.5$. Dash-dotted lines for the nonlinear case of dispersion $E = \pm\sqrt{c^2 p_z^2 + m_e^2 c^4}$ (parameter $\varepsilon = 0$).

Recall that system (9) corresponds to spin $\hbar/2$. In this case, all components of the wave function with indices 2 and 4 are zero. To obtain systems of equations corresponding to spin $-\hbar/2$, it is necessary to replace indices $1 \to 2$ and $3 \to 4$ in systems (9). Accordingly, by doing the same replacement of indices in solutions of system (9), we will get solutions for the case of spin $-\hbar/2$. In this case, all components of the wave function with indices 1 and 3 are equal to zero. Thus, eight linearly independent solutions are obtained corresponding to eight possible combinations of two branches of dispersion, two signs of total energy, and two possible directions of spin orientation, each in the form of four plane waves.

In the three-dimensional case, the generalized system of Dirac equations is expressed as follows:

$$\begin{cases} i\hbar \frac{\partial \Psi_1}{\partial t} = i\hbar c \left( -\frac{\partial \Psi_4}{\partial x} + i\frac{\partial \Psi_4}{\partial y} - \frac{\partial \Psi_3}{\partial z} \right) + \frac{m_e c^2}{\sqrt{1+\varepsilon^2}} (\Psi_1 - \Phi_1) \\ i\hbar \frac{\partial \Psi_2}{\partial t} = i\hbar c \left( -\frac{\partial \Psi_3}{\partial x} - i\frac{\partial \Psi_3}{\partial y} + \frac{\partial \Psi_4}{\partial z} \right) + \frac{m_e c^2}{\sqrt{1+\varepsilon^2}} (\Psi_2 - \Phi_2) \\ i\hbar \frac{\partial \Psi_3}{\partial t} = i\hbar c \left( -\frac{\partial \Psi_2}{\partial x} + i\frac{\partial \Psi_2}{\partial y} - \frac{\partial \Psi_1}{\partial z} \right) - \frac{m_e c^2}{\sqrt{1+\varepsilon^2}} (\Psi_3 - \Phi_3) \\ i\hbar \frac{\partial \Psi_4}{\partial t} = i\hbar c \left( -\frac{\partial \Psi_1}{\partial x} - i\frac{\partial \Psi_1}{\partial y} + \frac{\partial \Psi_2}{\partial z} \right) - \frac{m_e c^2}{\sqrt{1+\varepsilon^2}} (\Psi_4 - \Phi_4) \\ i\hbar \frac{\partial \Phi_1}{\partial t} = i\hbar c \left( -\frac{\partial \Phi_4}{\partial x} + i\frac{\partial \Phi_4}{\partial y} - \frac{\partial \Phi_3}{\partial z} \right) + \frac{m_f c^2}{\sqrt{1+\varepsilon^{-2}}} (\Phi_1 - \Psi_1)' \\ i\hbar \frac{\partial \Phi_2}{\partial t} = i\hbar c \left( -\frac{\partial \Phi_3}{\partial x} - i\frac{\partial \Phi_3}{\partial y} + \frac{\partial \Phi_4}{\partial z} \right) + \frac{m_f c^2}{\sqrt{1+\varepsilon^{-2}}} (\Phi_2 - \Psi_3) \\ i\hbar \frac{\partial \Phi_3}{\partial t} = i\hbar c \left( -\frac{\partial \Phi_2}{\partial x} + i\frac{\partial \Phi_2}{\partial y} - \frac{\partial \Phi_1}{\partial z} \right) - \frac{m_f c^2}{\sqrt{1+\varepsilon^{-2}}} (\Phi_3 - \Psi_3) \\ i\hbar \frac{\partial \Phi_4}{\partial t} = i\hbar c \left( -\frac{\partial \Phi_1}{\partial x} - i\frac{\partial \Phi_1}{\partial y} + \frac{\partial \Phi_2}{\partial z} \right) - \frac{m_f c^2}{\sqrt{1+\varepsilon^{-2}}} (\Phi_4 - \Psi_4) \end{cases} \quad (45)$$

In addition, we present the results obtained using the operator approach and Dirac notation. The Dirac equation for a free electron can be written in compact form:

$$i\hbar \frac{\partial}{\partial t} |\Psi_4\rangle = \widehat{H}_{D_4} |\Psi_4\rangle, \quad (46)$$

here

$$\widehat{H}_{D_4} = m_e c^2 \alpha_0 + c \sum_{j=1}^{3} \alpha_j \hat{p}_j. \quad (47)$$

the operator of the total energy (Hamiltonian) with the operators of the momentum component $\hat{p}_1 = \hat{p}_x = -i\hbar \frac{\partial}{\partial x}$, $\hat{p}_2 = \hat{p}_y = -i\hbar \frac{\partial}{\partial y}$, $\hat{p}_3 = \hat{p}_z = -i\hbar \frac{\partial}{\partial z}$; $|\Psi_4\rangle$ - four-component complex wave function; $\alpha_0, \alpha_1, \alpha_2, \alpha_3$ — matrices of size 4×4, called the Dirac alpha matrices:

$$\alpha_0 = \begin{pmatrix} I_2 & 0_2 \\ 0_2 & -I_2 \end{pmatrix}, \quad \alpha_1 = \begin{pmatrix} 0_2 & \sigma_x \\ \sigma_x & 0_2 \end{pmatrix}, \quad \alpha_2 = \begin{pmatrix} 0_2 & \sigma_y \\ \sigma_y & 0_2 \end{pmatrix}, \quad \alpha_3 = \begin{pmatrix} 0_2 & \sigma_z \\ \sigma_z & 0_2 \end{pmatrix}. \quad (48)$$

where $0_2$ and $I_2$ are zero and unit matrices of dimension 2×2, and $\sigma_j$ ($j = 1, 2, 3$) are Pauli matrices:

$$\sigma_x = \begin{pmatrix} 0 & 1 \\ 1 & 0 \end{pmatrix}, \quad \sigma_y = \begin{pmatrix} 0 & -i \\ i & 0 \end{pmatrix}, \quad \sigma_z = \begin{pmatrix} 1 & 0 \\ 0 & -1 \end{pmatrix}. \quad (49)$$

Each pair of alpha matrices is anticommuting, and the square of each is equal to one:

$$\alpha_j^2 = I_4, \quad j = 0, 1, 2, 3; \quad (50)$$

$$\alpha_i \alpha_j + \alpha_j \alpha_i = 0_4, \quad i, j = 0, 1, 2, 3 \ (i \neq j). \quad (51)$$

where $I_4$ and $0_4$ are zero and unit matrices of dimension 4×4. When squaring the Hamiltonian $\widehat{H}_{D_4}$ with taking into account the properties of the Dirac alpha matrices, we obtain:

$$\widehat{H}_{D_4}^2 = \left( m_e c^2 \alpha_0 + c \sum_{j=1}^{3} \alpha_j \hat{p}_j \right)^2 = m_e^2 c^4 \alpha_0^2 - c^2 \hbar^2 I_4 \Delta. \quad (52)$$

Consequently, after acting on the left side of the Dirac equation (46) by the operator $i\hbar\frac{\partial}{\partial t}$, and on the right side by the operator $\hat{H}_{D_4}$, we can obtain the four independent KGF equations:

$$-\hbar^2 \frac{\partial^2}{\partial t^2}|\Psi_4\rangle = \hat{H}_{D_4}^2|\Psi_4\rangle. \quad (53)$$

Accordingly, the generalized Dirac equation can also be written in compact form:

$$i\hbar\frac{\partial}{\partial t}|\Psi_8\rangle = \hat{H}_{D_8}|\Psi_8\rangle. \quad (54)$$

here

$$\hat{H}_{D_8} = \frac{m_f c^2}{\sqrt{1+\left(\frac{m_e}{m_f}\right)^2}} A_{0-} + \frac{m_e c^2}{\sqrt{1+\left(\frac{m_f}{m_e}\right)^2}} A_{0+} + c \sum_{j=1}^{3} A_j \hat{p}_j \quad . (55)$$

$|\Psi_8\rangle$ - an eight-component complex wave function; $A_{0-}$, $A_{0+}$, $A_1$, $A_2$, $A_3$ are matrices of size 4×4, which are a generalization of Dirac alpha matrices:

$$A_{0-} = \begin{pmatrix} 0_4 & 0_4 \\ -\alpha_0 & \alpha_0 \end{pmatrix}, A_{0+} = \begin{pmatrix} \alpha_0 & -\alpha_0 \\ 0_4 & 0_4 \end{pmatrix}, A_1 = \begin{pmatrix} \alpha_1 & 0_4 \\ 0_4 & \alpha_1 \end{pmatrix}, A_2 = \begin{pmatrix} \alpha_2 & 0_4 \\ 0_4 & \alpha_2 \end{pmatrix}, A_3 = \begin{pmatrix} \alpha_3 & 0_4 \\ 0_4 & \alpha_3 \end{pmatrix}. \quad (56)$$

For generalized Dirac alpha matrices, the following relations are fulfilled:

$$A_{0-}^2 = A_{0-}A_{0+} = \begin{pmatrix} 0_4 & 0_4 \\ -I_4 & I_4 \end{pmatrix}. \quad (57)$$

$$A_{0+}^2 = A_{0+}A_{0-} = \begin{pmatrix} I_4 & -I_4 \\ 0_4 & 0_4 \end{pmatrix}. \quad (58)$$

$$A_j^2 = I_8, \quad j = 1, 2, 3. \quad (59)$$

$$A_{0+}A_{0-} + A_{0-}A_{0+} = A_{0-}^2 + A_{0+}^2 = \begin{pmatrix} I_4 & -I_4 \\ -I_4 & I_4 \end{pmatrix}. \quad (60)$$

$$A_{0-}A_j + A_j A_{0-} = 0_8, \quad j = 1, 2, 3. \quad (61)$$

$$A_{0+}A_j + A_j A_{0+} = 0_8, \quad j = 1, 2, 3. \quad (62)$$

$$A_i A_j + A_j A_i = 0_8, \quad i, j = 1, 2, 3 \ (i \neq j). \quad (63)$$

where $0_8$ and $I_8$ are zero and unit matrices of dimension 8×8. When squaring the Hamiltonian $\hat{H}_{D_8}$ with taking into account the properties of generalized Dirac alpha matrices, we obtain:

$$\hat{H}_{D_8}^2 = \left( \frac{m_f c^2}{\sqrt{1+\left(\frac{m_e}{m_f}\right)^2}} A_{0-} + \frac{m_e c^2}{\sqrt{1+\left(\frac{m_f}{m_e}\right)^2}} A_{0+} + c \sum_{j=1}^{3} A_j \hat{p}_j \right)^2 = m_f^2 c^4 A_{0-}^2 + m_e^2 c^4 A_{0+}^2 - c^2\hbar^2 \Delta \quad . (64)$$

Accordingly, after acting on the left side of the generalized Dirac equation (54) by the operator $i\hbar\frac{\partial}{\partial t}$, and on the right side by the operator $\hat{H}_{D_8}$, the following equation is obtained:

$$-\hbar^2 \frac{\partial^2}{\partial t^2}|\Psi_8\rangle = \hat{H}_{D_8}^2|\Psi_8\rangle. \quad (65)$$

The last equation can be rewritten in the form of four independent systems of generalized KGF equations, with two equations in each, which are generalizations of the corresponding KGF equation.

## CONCLUSIONS

It is well known that the Dirac equation with a four-component wave function has only an optical branch of the dispersion relation. We have suggested the generalized Dirac equation with an eight-component wave function. This equation has both optical and acoustic branches of the dispersion relation, each of which is represented by branches with positive and negative energies. In the one-dimensional case, eight linearly independent solutions were obtained, each in the form of four plane waves. The solutions corresponded to eight possible combinations of two branches of dispersion, two signs of total energy, and two possible directions of spin orientation. In future work, we plan to discuss a number of issues related to the generalized Dirac equation with an eight-component wave function. These are charge density and current density, motion in a centrally symmetric field, and spin. We plan to investigate the nonrelativistic limit, which should give a generalization of the Schrödinger equation. It is necessary to give a physical interpretation of the physical nature of the new components $\Phi$ of the wave function. It is necessary to do this for both the optical and acoustic branches of the dispersion relation.

## ACKNOWLEDGMENTS

The authors express their gratitude to Oleg Markov, ScD, Evgeny Belkin, ScD, and Ilia Kopchinskii for discussions.

## CONFLICT OF INTEREST

The authors declare that they have no conflicts of interest.